\documentclass[onecolumn,technote]{IEEEtran}
\ifCLASSINFOpdf
\usepackage[pdftex]{graphicx}
\graphicspath{{../figures/}}
\DeclareGraphicsExtensions{.pdf,.jpeg,.png}
\else
\fi
\usepackage{array}

\usepackage[table,xcdraw]{xcolor}
\usepackage{multirow}

\usepackage{enumitem}
\usepackage[T1]{fontenc}

\usepackage[utf8]{inputenc} 

\begin{document}
%
\title{Implicit Theories and Self-efficacy in an Introductory Programming Course}


\author{\IEEEauthorblockN{F. Boray Tek \IEEEauthorrefmark{1},
Kristin S. Benli\IEEEauthorrefmark{1}, 
Ezgi Deveci\IEEEauthorrefmark{2} } 

\IEEEauthorblockA{\IEEEauthorrefmark{1}Department of Computer Engineering,
Işık University, İstanbul, Turkey}

\IEEEauthorblockA{\IEEEauthorrefmark{2}Departmenf of Physchology, Işık University, İstanbul, Turkey}

\thanks{Corresponding author: F. Boray Tek (email: boray.tek@isikun.edu.tr).}}

%



\maketitle

\begin{abstract}
Contribution: This study examined student effort and performance in an introductory programming course with respect to student-held implicit theories and self-efficacy.

Background: Implicit theories and self-efficacy shed a light into understanding academic success, which must be considered when developing effective learning strategies for programming.

Research Questions: Are implicit theories of intelligence and programming, and programming-efficacy related to each other and student success in programming? Is it possible to predict student course performance using a subset of these constructs?

Methodology: Two consecutive surveys (N=100 and N=81) were administered to non-CS engineering students in Işık University.

Findings: Implicit theories and self-beliefs are interrelated and correlated with effort, performance, and previous failures in the course and students explain failure in programming course with ``programming-aptitude is fixed'' theory, and also that programming is a difficult task for themselves.



\end{abstract}


\IEEEdisplaynontitleabstractindextext

%
\IEEEpeerreviewmaketitle

\section{Introduction}
%
%
%
%

\IEEEPARstart{L}{ar}ge numbers of today’s university students are required to enroll in introductory programming courses and significant percentages of these students struggle, repeat, and even switch programs due to difficulties with these courses. Consensus among university programming instructors is that programming is a very difficult task with a history of poor outcomes and no quick solutions \cite{ck09}, \cite{sg14}. University level programming courses generally have the highest dropout rates \cite{rrrh05}. Mostly, because that students who may have positive beliefs in their abilities across general subjects are frequently challenged with radical novelty in the course \cite{djik_89}, \cite{ccdos10}. 

Another factor is that programming is an area that requires motivation and practice, and above all, abstract thinking and a deep understanding, rather than surface memorization. Though Calculus also builds on similar skills, students are more acquainted with its problems, approach, language, and practice. On the other hand, programming requires studying a new language and a new approach to solving problems. In addition, the necessity of implementing and testing solutions using a computer produce many additional problems. Hence, for freshman students, programming requires harder work and deliberate practice more than other subjects.

On the other hand, modern psychology identifies new psychological constructs which shed a light into understanding academic success, which must be considered when developing effective learning strategies \cite{oak14}. Self-efficacy \cite{bnd94}, resilience, grit \cite{dpm07}, implicit theories (a.k.a mindset) \cite{dch95} are shown to affect student motivation and effort directly. Thus, academic achievement is not only the result of \emph{cognitive processes} -that acquire, organize, reuse formal information- but also strongly affected by \emph{motivational processes} that is related with expectation of outcomes for self-effort, and \emph{affective processes} that regulate emotional states and elicit emotional responses, e.g.\ stress or anxiety when faced with difficulty \cite{bnd94}.

Bandura \cite{bnd94} defines perceived self-efficacy as self-belief of a person in his/her capacity of accomplishing a \emph{specific} task at certain performance levels. Self-efficacy will determine how much effort a person will expend, how it confronts obstacles, and how resilient it will be during the course of that activity. The higher the self-efficacy the greater the effort, persistence, and resilience; which are central factors to learning success, more than previous experience \cite{rgdlb10}. 
Bandura emphasizes that a self-efficacy measure not specific to a task will be limited in explanatory and prediction power \cite{bnd06}. Thus, several self-efficacy scales were developed to measure subject-specific efficacy of students in different domains suchas general academic success, learning engineering \cite{mamaril16} , computer-use \cite{mrphy89}, mathematics \cite{betz83}, and programming \cite{rlw04}. 

On the other hand, several studies have incorporated Sherer's \cite{smm83} generalized self-efficacy scale to measure a person's self-belief of capacity to perform any novel or difficult tasks. Though it is not specific to a task, generalized self-efficacy scale is extensively used; validated and translated into 33 languages \cite{gse_url}) in clinical practice and behavioral analysis. However, Bandura does not support a task-independent efficacy scale, he states that self-efficacy in one field can affect another -even unrelated- field. Moreover, the degree of specificity in one context can be very different than another. In other words, a task can always be analyzed in different specificity levels, hence there can be two different self-efficacy scales which are task-specific, but one more general than the other. Thus, as Sherer \cite{smm83} and later Schwarzer \cite{schwrz95} conceptualized, a person can have a global belief in his/her capacity and performance in any \emph{novel or challenging task}. When it comes to succeeding in a foundational programming course, the question becomes whether a domain specific or the generalized self-efficacy measure is explanatory.
Hence, the first question of the current study is whether the generalized self-efficacy or programming-efficacy is more explanatory and predictive for student performance in an introductory programming course, and also whether two measurements are inter-related, or not? 

Along similar lines are the implicit theories \cite{dch95}, \cite{hcdlw99} that people hold about skills and abilities, and that play a role in how individuals explain the challenges that they are faced with in any learning situation. The implicit theories identified by Dweck and her colleagues are incremental and entity theories – more commonly referred to as fixed and growth mindsets. Entity theorists tend to believe that intellect is fixed. Their
goals are more performance-oriented, demonstrating competence and mastery. Exerting effort, perhaps through remediation, is considered a sign of weakness. As a result, entity theorists are more likely withdraw effort or procrastinate which will provide them excuses for underperformance.

Conversely, according to Dweck, incremental theorists are more oriented toward goals that reflect a concern with skill acquisition. A significant difference between these two approaches is that incremental theorists believe that intellect is not fixed and that increased effort leads to increased learning. Thus, they seek alternative paths – including remediation – to achieve the end-goal of acquiring knowledge or a skill. In other words, students with a growth mindset try to adapt themselves to radical novelty that they may sense in a programming course \cite{sg13}, which may also suggest that they have increased self-efficacy and resilience. In contrast, students with fixed mindsets have been shown to spend less time delving deeply into instructor feedback that may point them toward success in the course \cite{mblgd06}. Moreover, whereas a growth mindset may be strengthened, studies have reported increases in a fixed mindset over time as well e.g.\ \cite{ccdos10, sg14}. A student who believes that programming is a special talent, which may or not be linked to general intelligence, may easily feel a setback by the novelty and difficulty of programming. Dweck also identifies that implicit theories or mindset of a person can change from domain to domain. For example, a person can believe that English can be improved by studying hard, whilst believing mathematics or arts is only possible with a special talent that comes from genes. Therefore, mindset measurement scales were also introduced for different domains, such as programming \cite{ccdos10} \cite{sg13}, mathematics \cite{dwk08} and even politics \cite{dch95}.

In applying these concepts to the introductory programming context, the notions of fixed and growth mindsets, along with self-efficacy may be instrumental in helping to understand variation in student persistence and success in foundational courses. Connections have been suggested between other subjects (e.g. mathematics and science) and success in programming \cite{bl01}, as well as between academic self-efficacy and attitudes toward programming \cite{gsm12}, \cite{kkc09}. Some studies have shown that students may have high general self-efficacy or growth mindset, but they may struggle with domain-specific efficacy or mindset, such as programming \cite{ks12}, \cite{kork14}, \cite{sg13}, \cite{sg14}.

Ramalingham et al.\ \cite{rlw04} established a mutual relationship between programming-efficacy and performance in a programming course. Their work was adapted and applied to Turkish
engineering students in two different descriptive studies \cite{kork14}, \cite{ad09} which measured students’ programming-efficacy in a programming course based on C++ and Java, respectively. Scott and Ghinea \cite{sg14} have identified the connection between success and failure in introductory programming courses and students’ implicit theories of programming-aptitude.

The field examined in the present study is the intersection of implicit theories and self-efficacies, and their joint relation to success and failure in a university-level introductory programming course. The paper compares students' generalized theories about intelligence with domain-specific theories about programming aptitude; and generalized self-efficacies with programming-efficacies in their relation to desire to learn, effort, and performance (and failure). The study explores the following questions: Does student mindset for intelligence and programming are interrelated? Also does specific or general efficacies and mindsets are related? Is it possible to predict student performance using a subset of these measures?
At the other end, is it possible that the difficulty of programming and potential failure in the course leads to a belief of fixed programming aptitude and lower programming-efficacy? The latter discussion is enriched by comparing repeating students to first-timers. Repeating students beliefs and theories are particularly important because they may be suffering from lower self-efficacy and a fixed mindset from the previous experience. In addition, this mindset can be contagious for new students which will be somehow susceptible to repeating (and experienced) students' reviews on the course.

\section{Method}

The current study employs different scales to measure student implicit theories (hereafter mindsets) about 1) intelligence (generalized), 2) programming-aptitude (domain-specific); and different scales to measure 3) self-efficacy (generalized), 4) programming-efficacy (domain-specific). Performance is measured with several variables such as re-enrollment, course grade (and failure), general GPA (Grade Point Average). Student desire to learn is self-reported; whereas effort is measured by their self-reported study frequency. Correlations are examined to explore the relationship among different scales/variables; regression is performed to reveal whether student performance can be predicted; t-tests explored the belief differences among students who are first-timers and repeaters and who have failed and passed the course. The following questions and hypotheses will be tested:

H1- Student general intelligence mindset and general self-efficacy scores and domain-specific programming-aptitude and programming-efficacy scores are correlated with each other and with their effort and performance (course grade, repeats, GPA). Particularly, students who believe in improvable programming aptitude theory study more, perform better and express a positive desire to learn programming.

H2- Programming-aptitude and -efficacy scores can be used to predict student course performance.

H3- Students who fail the course tend to associate failure with fixed programming aptitude; and have lower self-efficacy. Moreover, re-enrolled students believe in fixed programming aptitude theory more than the first-time enrolling students.

H4- As the course progresses, their mindsets will shift from growth to fixed; and efficacies will decrease.



\subsection{Course and Data Collection}\label{course}

Introduction to Programming (CSE-101) is a mandatory course for non-CS engineering and IT majors at I\c{s}{\i}k University. The objective of this course is to introduce the basic concepts of programming, and encourage solving problems by computer programming. The 14-week course introduces the fundamentals of programming, where only the basic constructs $\lbrace variables, selections, loops, arrays, methods \rbrace$ are introduced but object-oriented concepts (e.g.\ classes, instances) are left out. In addition to a three-hour per week lecture session where programming constructs are introduced, and problems and algorithms are discussed/solved, a laboratory session (two hours per week) introduces small problems/solutions to provide hands-on experience on the Java language platform. The course performance and final grade are calculated over 3x Exams (80\%), 5x Quizzes (10\%), 12 weeks of laboratory attendance\&practice (10\%). Exams had 4 or 5 programming questions focusing in different constructs and problem-solving strategies. Quizzes were the open-book problems that students solved in pairs, during lectures. Laboratory sessions included usually four to five solved problems, and two more to be solved by the students. Attending and trying to solve problems in the laboratory was sufficient to get laboratory grade for the week.

The study was conducted in Spring 2016 in two parts: one in the 8th week (pre-survey), and the other in the 13th week (post-survey) of the semester. The surveys were prepared in Turkish (the course is taught in English); created with Google Docs; then filled out online by the students who attended laboratory sessions. The pre-survey was administered one week before the first midterm exam and the post-survey as conducted approximately two weeks before the final exam. Before the pre-survey, the course had completed topics until nested loops, whereas methods and arrays were completed in post-survey, only the basic sorting algorithms were left before the post-test. Prior to the pre-survey, students have already seen lots of mini programming problems, and so they have already faced with the challenge. Prior to the post-survey, the students had received 60\% of the course grades. Hence, they were almost able to predict their performance. To evaluate student self-beliefs and theories it is important that they have some experience and intuition about the task and challenge, however, still, it could be useful if the study included another test in the first week of the semester as it was also pointed by the participant data distribution (see the next section).

\subsection{Participants}

Table \ref{demographicInfo} summarizes some demographic information about the participants. In total, there were 193 Turkish students in three different course sections. The pre- (first row) and post- (second row) surveys were completed by 100 and 81 students, respectively. 65 participants completed both the pre- and post-surveys (third column). The participants were mostly male and the vast majority of the respondents were engineering students, excluding computer and software majors.

\begin{figure}
	\centering
	\includegraphics[width=0.5\linewidth]{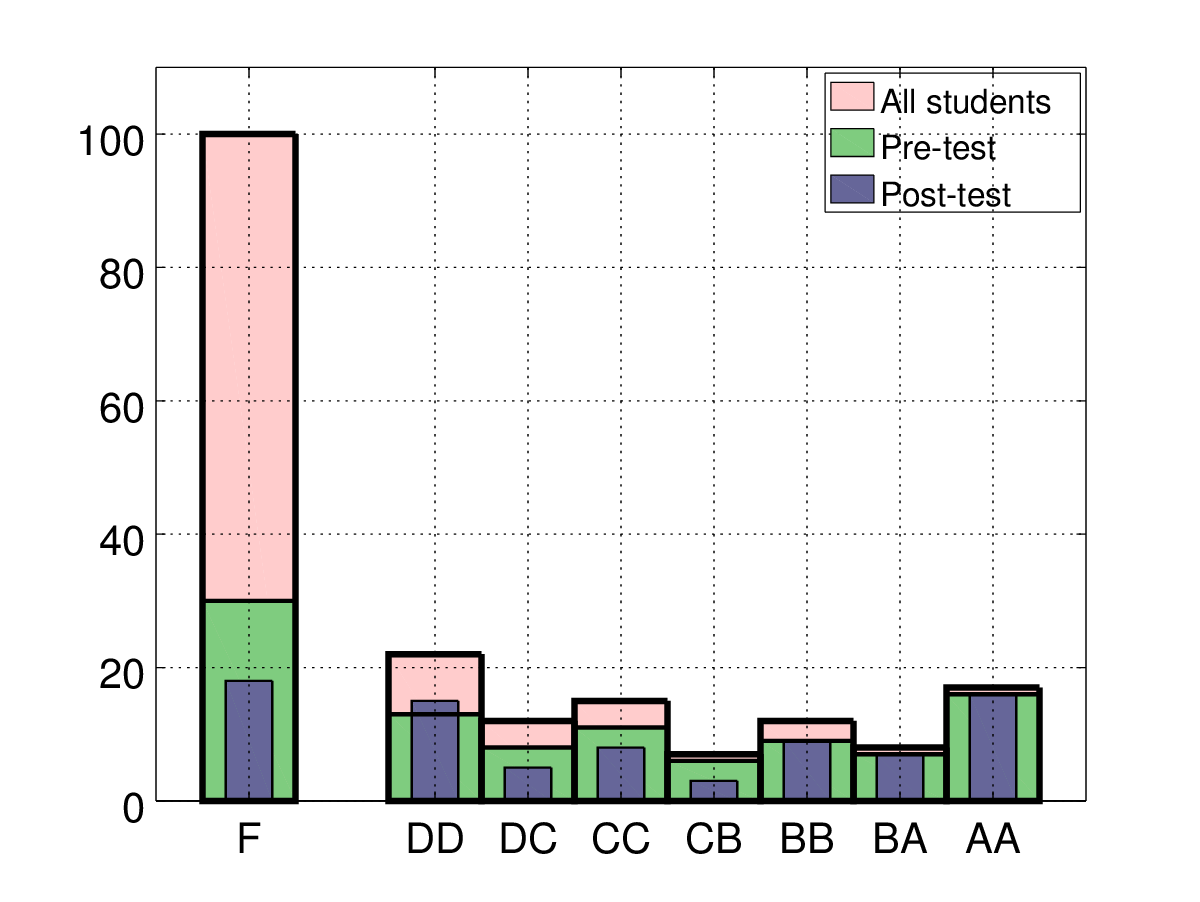}
	\caption{Letter grade distribution of (pink-light) all CSE101 (Spring 2016) students, (green-medium) respondents of the pre-survey, (blue-dark) respondents of the post-survey.
		}
	\label{fig:plotpregradesdistribution}
\end{figure}

More than half of the respondents were taking a programming course for the first time. The remainder had failed and re-enrolled, not because they insist on learning programming, rather because the course is mandatory. Figure \ref{fig:plotpregradesdistribution} shows the grade distribution among the participating students of the pre and post surveys versus all students. While the majority of the failed students did not participate -perhaps because of the late application of the surveys-, the participants of both surveys had balanced distributions among different performance levels.

\begin{table*}[ht]
\centering
\caption{Demographic information}
\label{demographicInfo}
\begin{tabular}{| l | c | l  r | l  r | l  r | l  r | l r | c |}
\rowcolor[HTML]{9B9B9B} 
\hline
Survey & Total &  \multicolumn{2}{c}{Gender} & \multicolumn{2}{c}{Age} & \multicolumn{2}{c}{Department} & \multicolumn{2}{c}{\#Repeats} & \multicolumn{2}{c}{Course Grade} & Avg.\\
\rowcolor[HTML]{9B9B9B} 
{}   & Participants & Female   & Male    & Avg.   & Max  & Eng. & IT & First-Timer& Repeater & Pass & Fail  & GPA\\ \hline
\rowcolor[HTML]{EFEFEF} 
Pre   & 100 &  39.0\% & 61.0\%   & 20.81  & 27 & 93.0\% & 7.0\% & 62.0\% & 38.0\% & 70.0\% & 30.0\% & 1.73 \\	 \hline
Post   & 81 &  33.3\% & 66.6\%   & 20.90  & 29 & 88.8\% & 11.1\% & 66.6\% & 33.3\% & 77.7\% & 22.2\% & 1.92 \\	\hline
\rowcolor[HTML]{EFEFEF} 
Pre\&Post   & 65 & 38.5\%  &  61.5\%   & 20.75  & 27 & 92.3\% & 7.7\% & 69.2\% & 30.8\% & 81.5\% & 18.5\% & 1.99 \\	\hline
\end{tabular}
\end{table*}

\subsection{Scales and Measures}
The scales and observed variables are explained below. Readers can find the complete survey, its English translation, and a factor analysis of newly adapted scales online (see Supplementary material). 

\subsubsection{Mindset for Intelligence \textbf{(MNDINT)}}
The first part of the survey examined students' mindsets about intelligence using a subset of Dweck's scale \cite{dch95}. Four items were taken from the original scale and translated into Turkish. Then they were back-translated by a translator and checked by another native English instructor. Two of the items supported the belief in a fixed intelligence theory, e.g. \emph{``People have a certain amount of intelligence, and they can’t really do much to change it''}.  Whereas two others supported the belief in a growth intelligence, e.g.\  \emph{``No matter which intelligence level he or she has, a person can change it even if just a bit.''}. These questions were scored on a six-point scale as in the original. The principal components analysis (PCA) revealed that the four items were loading to only one factor between $.66$ and $.85$ (explained variance $= 58.85$). The literature \cite{dmb14} has different findings about the number of factor loadings of mindset scale, perhaps due to the use of different subsets. Our results support Dweck's \cite{dch95}. Testing the reliability, Cronbach’s alpha value was $.76$; and item-total correlations were between $.71$ and $.82$. Cr. alpha is a commonly used measure in the analysis of homogeneity and unidimensionality of scales of more than one items. However, the reliability of Cr. alpha is known to be adversely affected by the number of items in a test \cite{td11}. The suitability of the data for factor analysis can be examined by the Kaiser-Meyer-Olkin (KMO) partial correlation coefficient and the Barlett’s test of sphericity. The KMO coefficient gives information about whether the data matrix is suitable for factor analysis and whether the data structure is suitable for factoring. For factorability, it is expected that KMO will be higher than .60. The Barlett’s test examines whether the test items are correlated with themselves, and have some level of correlation with other items. The Kaiser-Meyer-Olkin (KMO) measure of sampling adequacy was $.65$, and Bartlett’s test of sphericity was significant ($\chi^2(6)=105.39, \, p<.001$).

\subsubsection{Mindset for Programming Aptitude \textbf{(MNDPRG)}}
Following Scott and Ghinea \cite{sg14} the authors adapted Dweck's mindset scale to the programming context to measure whether students believed: programming skills can be improved by practice; or it is a fixed skill. Simply, the word \emph{``intelligence''} in the intelligence mindset scale was replaced with the words ''programming-aptitude''. For example, the first question was, \emph{``People have a certain level of \emph{programming-aptitude}, and they can’t really do much to change it''}.  Again, items were scored on a six-point scale. PCA revealed that the four items were loading to one factor between $.66$ and $.79$ (explained variance = $55.61$). Cronbach’s alpha value was $.73$; and item-total correlations changed between $.70$ and $.78$. The KMO measure of sampling adequacy was $.61$; and Bartlett’s test of sphericity was significant ($\chi^2(6)=87.69, \, p<.001$).

\subsubsection{Self-efficacy \textbf{(EFCGEN)}}
To construct and analyze (generalized vs specific) correspondence, as of the general intelligence mindset scale and the programming aptitude, the current study employed the generalized self-efficacy scale from Sherer \cite{smm83} which was adapted to Turkish by Y{\i}ld{\i}r{\i}m \cite{yi10}. The survey consists of 17 questions that are scored on a five-point scale. The set includes both positive and negative questions such as \emph{``I avoid facing difficulties''(-), ``I give up easily''(-), ``Failure increases my perseverance''(+)}. The scale is more general than the academic self-efficacy employed in relevant studies \cite{kkc09,kork14}. Cronbach's alpha was $.89$.

\subsubsection{Programming-efficacy \textbf{(EFCPRG)}}
There are existing adaptations of self-efficacy to the domain, such as the programming-efficacy scales that were adapted to C and Java language platforms \cite{ad09,kork14}. These scales are layered, and rate student self-efficacy in using different level programming constructs and solving problems of increasing difficulty. However, because the CSE101 course is not object-oriented, the authors could not utilize these programming-efficacy surveys directly. Moreover, a general vs. specific comparison required measuring students' programming beliefs in one-to-one correspondence with their general self-efficacy. Therefore, the authors adapted the latter to measure programming-efficacy from a self-assessment of difficulty perspective. Starting with a subset of the original scale (EFCGEN), simply the keyword programming was inserted in the appropriate positions to form six questions that were scored on a five-point scale. Items included both positive and negative questions such as \emph{``I am sure that I can learn programming''(+), ``Programming is a difficult task for me''(-), ``If a programming problem seems too complicated, I do not try solving it''(-)}. PCA revealed that the six items were loading to one factor between .69 and .85 (explained variance $=60.03$). Cronbach’s alpha was $.86$, and item-total correlations were between $.73$ and $.83$. The KMO measure of sampling adequacy was $.81$, and Bartlett’s test of sphericity was significant ($\chi^2(15)=233.22, p < .001$).

\subsubsection{Effort\&Practice \textbf{(EFT)}}
To measure student effort, the survey included three items on a six-point scale which questioned self-reported programming effort. The set includes \emph{"I learn programming by following and noting programs in lectures/labs", "I learn programming by coding lecture/lab problems.", "I learn programming by writing programs on a computer myself"}. PCA revealed that the three items were loading to one factor .61 to .81 (explained variance $=54.6$). Cronbach's alpha was $.72$. The KMO measure was $.80$, and Bartlett’s test was significant ($\chi^2(6)=61.13, p < .001$).

\subsection{Single Observed Variables}
\subsubsection{Course repeats \textbf{(REP)}} indicates the number of previous enrollments in the course (survey question 35 in the supplementary material). More than one-third of the respondents had previously failed and re-enrolled in this mandatory course.

\subsubsection{Course grade \textbf{(GRD)}} is the end-of-term performance grade of each respondent (\{F=0, DD=1.0, DC=1.5, CC=2.0, CB=2.5, BB=3, BA=3.5, AA=4.0\}), calculated from the average described in Section \ref{course}. Students receiving a grade below CC are advised to repeat, whereas students with an F must repeat the course. 

\subsubsection{Laboratory Attendance \textbf{(ATT)}} rates the student attendance to laboratory sessions from 0 to 10. This is also included in the course grade.

\subsubsection{Cumulative Grade Point Average \textbf{(GPA)}} is over all courses that the student has taken, as recorded on their transcript. Usually, Engineering students enroll for CSE101 in their first year, the second term. However, as students can fail and re-enroll in different terms, the effect of CSE101 on the GPA may be high or low, but not zero.

\subsubsection{Desire to learn \textbf{(DES)}} rates student's answer to the question \emph{``How much do you want to learn programming?''} (question 36, only in post-survey). Respondents answered on a five-point scale; two-thirds expressed that they very enthusiastic to learn programming.

\subsubsection{Intelligence proportional to programming aptitude \textbf{(INT$\propto$PRG)}} rates agreement with the statement \emph{``intelligence and programming aptitude are proportional''}, (question 37) on a six-point Likert scale. Mean of pre- and post-surveys denote that on average students slightly disagree with the proposition.

\section{Data Analysis}
\begin{table*}[t]

\centering
\caption{Correlations between scales and single observed variables}
\label{tabcorl}
\resizebox{\linewidth}{!}{%
\begin{tabular}{llrlllllllllllll}
\hline
& \textbf{Rng} &\textbf{Pr/Ps} &  \textbf{Mean} & \textbf{SD} & \textbf{MNDINT} & \textbf{MNDPRG} & \textbf{EFCGEN} & \textbf{EFCPRG} & \textbf{EFT} & \textbf{REP} & \textbf{GRD} & \textbf{ATT} & \textbf{GPA} &\textbf{DES}&\textbf{INT}$\propto$\textbf{PRG}\\ 

\hline

\rowcolor[HTML]{EFEFEF}
\textbf{MNDINT}& $_{\left[1-6\right]}$   & \textbf{Pr}   & 3.62      & 1.04        & -           & 0.22* & 0.29** & 0.14 & 0.02 & -0.21* & -0.08 & -0.11 & -0.03 & - & 0.06\\ 
\rowcolor[HTML]{EFEFEF} 
\textbf{MNDINT}& $_{\left[1-6\right]}$   & \textbf{Ps}   & 3.49       & 1.0        & -           & 0.31** & 0.11 & 0.16 & 0.16 & -0.21 & 0.15 & 0.02 & 0.21 & -0.08 & -0.14 \\ \hline

\textbf{MNDPRG}&$_{\left[1-6\right]}$   & \textbf{Pr} & 4.48      & 0.94      &             & -           & 0.32** & 0.66*** & 0.53*** & -0.28** & 0.20* & -0.07 & 0.16 & - & 0.03	\\
\textbf{MNDPRG}&$_{\left[1-6\right]}$   & \textbf{Ps} & 4.11       & .99       &             & -           & 0.29* & 0.65*** & 0.44*** & -0.33** & 0.48*** & 0.01 & 0.36*** & 0.33** & 0.02	\\ \hline

\rowcolor[HTML]{EFEFEF}
\textbf{EFCGEN}&$_{\left[1-5\right]}$   & \textbf{Pr} & 3.73      & 0.54      &             &             & -           & 0.38*** & 0.20* & -0.20* & 0.19 & 0.06 & 0.23* & - & 0.0	\\
\rowcolor[HTML]{EFEFEF} 
\textbf{EFCGEN}&$_{\left[1-5\right]}$   & \textbf{Ps} & 3.62       & 0.61       &             &             & -           & 0.46*** & 0.20 & -0.17 & 0.21 & 0.16 & 0.31** & 0.37*** & 0.09	\\ \hline
\textbf{EFCPRG}&$_{\left[1-5\right]}$   & \textbf{Pr} & 3.66       & 0.93        &             &             &             & -           & 0.61*** & -0.26* & 0.25* & -0.13 & 0.27* & - & 0.16	\\
\textbf{EFCPRG}&$_{\left[1-5\right]}$   & \textbf{Ps} & 3.58       & 0.79        &             &             &             & -           & 0.50*** & -0.36*** & 0.52*** & -0.11 & 0.45*** & 0.44*** & 0.24*\\ \hline
\rowcolor[HTML]{EFEFEF}
\textbf{EFT}&$_{\left[1-6\right]}$   & \textbf{Pr} & 4.35      & 0.54      &       &   &   &             & -           & -0.11 & 0.15 & -0.01 & 0.07 & - & 0.18	\\
\rowcolor[HTML]{EFEFEF} 
\textbf{EFT}&$_{\left[1-6\right]}$   & \textbf{Ps} & 4.26       & 1.03      &       & &     &             & -           & -0.13 & 0.24* & -0.04 & 0.25* & 0.22* & 0.33** 	\\
\hline
\textbf{REP}&$_{\left[1+\right]}$   & \textbf{Pr} & 1.79       & 1.21        &     &        &             &             &             & -           & -0.16 & 0.11 & -0.34*** & - & -0.04\\

\textbf{REP}&$_{\left[1+\right]}$   & \textbf{Ps} & 1.65        & 1.12      &  &             &             &             &             & -           & -0.32** & -0.05 & -0.40*** & -0.20 & -0.18 \\ \hline
\rowcolor[HTML]{EFEFEF}
\textbf{GRD}&$_{\left[0-4\right]}$  & \textbf{Pr} & 1.78       & 1.49       &           &             &             &             &             & - &  -         &   -0.05 & 0.79*** & - & -0.01	\\
\rowcolor[HTML]{EFEFEF}
\textbf{GRD}&$_{\left[0-4\right]}$  & \textbf{Ps} & 1.99        & 1.49      &  &             &             &             &             &             & -           & -0.14 & 0.78*** & 0.32** & 0.25*	\\ \hline
\textbf{ATT}&$_{\left[0-10\right]}$  & \textbf{Pr} &   4.9   &  3.2       &            &             &             &            &             &             &            &    -   & -0.04 & - & 0.06   \\

\textbf{ATT}&$_{\left[0-10\right]}$   & \textbf{Ps} & 4.48        & 3.23       & &            &             &             &             &             &             & -           & -0.08 & 0.34** & 0.18 \\ \hline
\rowcolor[HTML]{EFEFEF}
\textbf{GPA}&$_{\left[0-4\right]}$   & \textbf{Pr} & 1.73        & 1.44        & &            &             &             &             &             &             &             & -           &  - & 0.07	\\
\rowcolor[HTML]{EFEFEF}
\textbf{GPA}&$_{\left[0-4\right]}$   & \textbf{Ps} & 1.90        & 0.82        &  &           &             &             &             &             &             &             & -           &  0.20 & 0.19	\\ \hline

\textbf{DES}&$_{\left[1-6\right]}$   & \textbf{Pr} & -        & -       &    &         &             &             &             &             &             &             &             & -           & -	\\

\textbf{DES}&$_{\left[1-6\right]}$   & \textbf{Ps} & 3.63        & 1.27        &    &         &             &             &             &             &             &             &             & -           &0.22*\\ \hline
\rowcolor[HTML]{EFEFEF}
\textbf{INT}$\propto$\textbf{PRG}&$_{\left[1-6\right]}$   & \textbf{Pr} & 3.27        & 1.41      &       &      &             &             &             &             &             &             &             &             &-	\\
\rowcolor[HTML]{EFEFEF}
\textbf{INT}$\propto$\textbf{PRG} &$_{\left[1-6\right]}$   & \textbf{Ps} & 3.47        & 1.43        &      &       &             &             &             &             &             &             &             &             &-	\\ \hline
\hline

\multicolumn{16}{l}{\footnotesize{Note: Rng: range, Pr: Pre, Ps: Post survey; MNDINT: Mindset intelligence; MNDPRG: Mindset Programming-aptitude; EFCGEN: General self-efficacy; EFCPRG: Programming-efficacy; EFT: Effort}} \\
\multicolumn{16}{l}{\footnotesize{REP: Course repeats, GRD: Course grade, ATT: Laboratory attendance GPA: Cumulative GPA, DES: Desire to learn programming, INT$\propto$PRG}: Intelligence proportional to programming} \\
\multicolumn{16}{l}{N$_{Pr}$=100, N$_{Ps}$=81; $*p<.05$; **$p<.005$; ***$p<.001$}
\end{tabular}%
}
\end{table*}

\subsubsection{\textbf{Hypothesis 1 (Correlations): Do Student general intelligence mindset and general self-efficacy scores and domain-specific programming-aptitude and programming-efficacy scores are correlated with each other and with their effort and performance (course grade, repeats, GPA). Do students who believe in improvable programming aptitude theory study more, perform better, and express a positive desire to learn programming?}} 
Table \ref{tabcorl} shows data from pre- (Pr) and post-surveys (Ps) with means (MN) and standard deviations (SD). The scores were normed with respect to the number of questions in each survey. 
The domain-specific scales: the mindset for programming (MNDPRG), programming efficacy (EFCPRG) and effort (EFT) are all positively inter-correlated; and also positively correlated with course grade and negatively correlated with course repeats. The desire to learn programming (DES) is positively correlated with the mindset for programming, general self-efficacy and more strongly with programming efficacy.
The stronger correlations in pre-survey generally agree with the post surveys'. The results reveal that student who believe in an improvable programming aptitude (higher in MINDPRG) have higher programming efficacy. They also study more, get higher grades in the course; less likely to repeat the course; and express a positive desire for learning programming. An observation is that effort and the agreement for "intelligence and programming aptitude is proportional" statement (INT$\propto$PRG) is significantly positively correlated. The scatter plot of the data (not shown) showed that the majority of the students report effort above average, whereas only a few completely disagree with "intelligence and programming aptitude" are proportional.

\subsubsection{\textbf{Hypothesis 2 (Performance prediction): Can programming aptitude and programming-efficacy scores (domain-specific scales) be used to predict student course performance (grade)?}}

A multiple linear regression analysis was conducted to predict student course grade by the pre-survey scale scores. Table \ref{tabreg} shows that none of the predictors had significant contribution to the model, despite a weak regression ($F$(4;95)=2.6, $p$=.041, R$^2$=.1, R$^2_{Adj}$=.06). Thus, grade prediction cannot be confirmed.



\begin{table}[tb]
	\begin{small}
		\centering
		\caption{Multiple Linear Regression Analysis of Grade}
		\label{tabreg}
		\begin{tabular}{llllll}
			\rowcolor[HTML]{C0C0C0}
			\multicolumn{6}{l}{\cellcolor[HTML]{C0C0C0}Model 1 (Course grade)}     \\
			\rowcolor[HTML]{EFEFEF}
		   \textbf{Pre Survey} & B     & SE(B) & $\beta$ & t & Sig.(p) \\

			\textbf{MNDINT}                                  & -.47  & .30  & -.16   & -1.58 & .117 \\
			\rowcolor[HTML]{EFEFEF} 
			\textbf{MNDPRG}                                  & .23  &  .42& .07 & .55 & .578 \\
			
			\textbf{EFCGEN}                                  & .79 &  .60 & .14 & 1.31 & .193 \\
			\rowcolor[HTML]{EFEFEF} 
			\textbf{EFCPRG}                                  & .57  &  .44 & .17 & 1.30 & .196 \\

			\multicolumn{6}{l} {\textbf{$R^2=.1$}, $R^2_{adj}=.06$} 	
			\\
			\end{tabular}
			
		\vspace{-0.5cm}
		
%
\end{small}
	\end{table}

\subsubsection{\textbf{Hypothesis 3 (Previous fails and re-enrollment): Do students who fail the course tend to associate failure with fixed programming aptitude, and have lower self-efficacy? Do previously failed (and re-enrolled) students believe in fixed programming aptitude theory more than the first-time enrolling students?}}

Table \ref{tabttest} shows independent-samples t-tests (from pre-survey N$_{100}$) that compare MNDINT, MNDPRG, EFCGEN and EFCPRG scores for the first-timer and re-enrolled students. There was a significant difference in MNDPRG and EFCPRG scores (respectively, t$_{(79)}$=3.58, $p$<.001; t$_{(98)}$=3.137, $p$<.001). The lower values of programing mindset of re-enrolled students suggest that a past failure amplified their belief on fixed programming aptitude. The post-survey data shows that the difference was larger at the end of the term. In addition, t-tests from post-survey show that views of students who fail and pass differed significantly regarding mindset for programming, self- and programming efficacy: MNDPRG (t$_{(79)}$=-2.1, $p$<.001), EFCGEN (t$_{(79)}$=-4.45, $p$<.05) and EFCPRG (t$_{(79)}$=-5.54,  $p$<.001), respectively.

\subsubsection{\textbf{Hypothesis 4 (Change during the course): Does student mindset shift from growth to fixed; and efficacies decrease during the course.}} 
Table \ref{tabcorl} shows that scores for all five scales dropped from pre-survey to the post-survey. For the students who have attended both pre and post-survey, a paired-samples t-test was conducted to compare MNDINT, MNDPRG, EFCGEN and EFCPRG scores (N$_{65}$). Though, all the mean values dropped from the pre- to post-survey, the only significant change was for MNDPRG (M$_{pre}$=4.46, SD$_{pre}$=.97 and M$_{post}$=4.2, SD$_{post}$=.96; t$_{(64)}$=2.5; $p$=.015;) and marginally for EFCGEN( M$_{pre}$=3.75, SD$_{pre}$=.6 and M$_{post}$=3.59, SD$_{post}$=.67; t$_{(64)}$=2.4, $p$=.052). 


\begin{table*}[tbh]

\centering
\caption{Independent Group T-Tests Between First Time and Re-enrolling Students; Failed and Passed Students}
\label{tabttest}
\begin{tabular}{llllllllllllll}
\rowcolor[HTML]{C0C0C0} 
                                               &                                                              & \multicolumn{3}{l}{\cellcolor[HTML]{C0C0C0}MNDINT} &   \multicolumn{3}{l}{\cellcolor[HTML]{C0C0C0}MNDPRG} &   \multicolumn{3}{l}{\cellcolor[HTML]{C0C0C0}EFCGEN} &   \multicolumn{3}{l}{\cellcolor[HTML]{C0C0C0}EFCPRG} \\
\rowcolor[HTML]{EFEFEF} 
                                               &                                                              & M        & SD      & t                         &   M       & SD     & t                           &   M       & SD     & t                           &   M       & SD     & t                           \\
\cellcolor[HTML]{C0C0C0}Pre-enrollment & \cellcolor[HTML]{EFEFEF}First-timer (n=62) & 3.71    & 0.93    &                           &   4.74   & 0.81   &                             &   3.81   & 0.45  &                             &   3.84   & 0.83   &                             \\
\cellcolor[HTML]{C0C0C0}{Pre-survey N=100}          & \cellcolor[HTML]{EFEFEF}Re-enrolled (n=38)         & 3.49    & 1.19    & \multirow{-2}{*}{1.023}    &   4.06   & 0.99   & \multirow{-2}{*}{3.580***}   &   3.62   & 0.66   & \multirow{-2}{*}{1.614}      &   3.26   & 0.98   & \multirow{-2}{*}{3.137**}    \\

\multicolumn{14}{l}{}                \\

\cellcolor[HTML]{C0C0C0}Performance     & \cellcolor[HTML]{EFEFEF}Failed  (n=46)               & 3.42    & 0.99    &                           &   3.7   & 0.92   &                             &   3.46   & 0.71  &                             &   3.22   & 0.67   &                             \\
\cellcolor[HTML]{C0C0C0}Post-survey N=81                & \cellcolor[HTML]{EFEFEF}Passed (n=35)               & 3.59    & 1.02    & \multirow{-2}{*}{-.741}   &   4.6   & 0.88   & \multirow{-2}{*}{-2.101*}    &   3.76   & 0.53   & \multirow{-2}{*}{-4.45***}  &   4.06   & 0.68   & \multirow{-2}{*}{-5.54***} \\

\multicolumn{14}{l}{\footnotesize{*$p<.05$, **$p<.01$, ***$p<.001$. }}{}
\end{tabular}
\end{table*}

\section{Discussion}
Mindset for programming aptitude (MNDPRG) represents the belief in improvable programming aptitude, whereas programming-efficacy (EFCPRG) is student's self-belief of his/her capacity to learn programming,  or in other words, his/her assessment of the challenge of programming from his/her self-perceived capacity. The generalized scales represent the belief in improvable intelligence (MNDINT), and task-independent self-efficacy (EFCGEN), respectively. The correlation analysis of data from surveys has several implications. First, as expected, domain-specific measures MNDPRG and EFCPRG are more relevant to programming than generic scales. The generic scales had some correlation with their domain-specific counterparts. Second, MNDPRG and EFCPRG are highly correlated. Therefore student's belief in whether programming aptitude can be improved by effort is somehow related to his/her belief in self-capacity to learn programming, and vice-versa. Furthermore, both measures correlate significantly with self-reported effort (EFT). Hence, for a significant portion of the survey students, a growth mindset for programming and higher programming-efficacy means higher effort. However, the weak correlation between self-reported effort and course grade shows that not all hard-working students get higher grades (GRD), despite the positive correlation between GRD, MNDPRG and EFCPRG. This may be the reason for the correlation among (for some of) the students who self-report higher effort\&practice but also agree with "programming aptitude and intelligence is proportional". Confirming Scott and Ghinea's \cite{sg14}, effort\&practice is very related to mindset for programming. The current paper shows that both are related with programming-efficacy. 

The analysis also revealed a correlation between self-expressed desire to learn (DES) and programming mindset and particularly with programming-efficacy. Furthermore, data supports that students who believe in improvable programming aptitude theory express much desire to learn programming, study more, and get higher course grades; and they think programming is a manageable challenge for themselves. Since both mindset and efficacy are related with goal settings, this correlation among programming efficacy and desire may also indicate a self-revision of goals in the face of difficulty, such that "Programming is difficult for me, I do not want to learn programming!". Students who declared less desire also had lower programming efficacy and are aligned towards fixed programming aptitude theory. However, whether these correlations indicate cause and effect or attribution relation can be answered after discussing other results.

Seeing the correlations, what is the possibility of predicting the course grade with mindset and efficacy measures. The regression analysis in the pre-survey has shown prediction would not be significant, at least for all students. This may have proven the obvious fact that course grade is a complex outcome that depends on many factors. However, it may still be possible to explore prediction for a subset of students, such as repeaters, which was not explored here. The lack of predictive power of implicit theories in course grade or GPA, align with Kornilova's \cite{kkc09} findings that implicit theories of intelligence had no predictive power on GPA. However, as Dweck shows \cite{dwk08} the effect on achievement may reveal itself over time as it acts indirectly by impacting goal orientations. 

The decreasing self- and programming-efficacies and decreasing belief in an improvable programming aptitude show that all students of the course (sampled by the participants) revise their theories and self-beliefs negatively upon facing the challenges, which is parallel with the attribution theory and relevant research \cite{ccdos10}, \cite{sg14}, \cite{kkc09}.

Programming mindset and efficacy scores of students who have failed the course were significantly lower than the successful ones. Therefore the strong correlations of these two factors with the course grade, but lack of significant prediction result, supports the attribution theory more than a predictive capacity. In other words, students who had trouble coping with the programming challenge and foreseen an upcoming failure, attribute it to the limited aptitude and also to that programming is a difficult task for themselves.

Hong et.\ al.\ \cite{hcdlw99} proposed that implicit theories can set up frameworks which can be filled with attributions in the case of failure. In the current case, these attributions may feed the fixed programming aptitude theory in a cyclic way, and contribute to future failures for repeating students. The significantly lower MNDPRG and EFCPRG scores of re-enrolled students support this conclusion. Thus, some of the prospective students for the next term would have pre-attributed the failure to their fixed programming aptitude.

There are several limitations of this study which may infer external validation of its results. First of all, while sample sizes seem sufficient for statistical analysis, the participants were from the same institution. Second, a major portion of the failed students did not even participate in the first survey. So our results may be biased to reflect beliefs of persistent students. Third, some of the scales such as effort relied only on self-reported values to indicate study behavior. An objective measure such as externally counted "number of programs written, number of questions attempted" could be more objective.
Gore \cite{gore06} showed that predictive power of efficacy depend on when efficacy beliefs are measured and types of efficacy questions. Hence, one can argue that the application of the pre-survey on the 8th week (before the first midterm) may have affected measurements of programming mindset and efficacy, since students may have already foreseen their own performance. However, to measure self-efficacies (in a specific task), the person must be exposed to the task. 
Therefore this study had delayed application of the pre-survey also that did not employ detailed programming efficacy questionnaires such as developed in \cite{rlw04,ad09}, which included questions such as "I understand OOP or Java Applets".

\section{Conclusion}

The comparison of implicit theories of intelligence, programming aptitude and self-efficacy in relation to student effort and performance in programming has confirmed and also enriched some of the earlier findings \cite{sg14} that students develop domain-specific implicit theories and self-beliefs. 
In particular, the current study shows that students who believe in improvable programming aptitude and have higher programming-efficacy study more, and get higher grades. On the other side, students who are re-enrolled to the course support a fixed programming aptitude theory more and have lower programming efficacy. Thus they come pre-attributed the previous failure(s) to ''programming is a difficult task for me, related to the pre-acquired skills/genes more than effort''. Therefore, belief in a fixed programming-aptitude and having a low programming-efficacy significantly increases the likelihood of course fails and thus repeats. Fortunately, it is possible to detect some of those students with the measures employed in this study. Therefore, an intervention must target increasing programming-efficacy and the false belief that ``the programming aptitude is fixed''; and show to the student that programming skill is improvable by practice. Acknowledging students with the results that the failure/repeats and the fixed programming-aptitude and low programming-efficacy is correlated, can be the first step as an intervention to strengthen the belief in a growth programming aptitude. Bandura \cite{bnd06} emphasizes that mastery experience is the most effective source of self-efficacy. Hence, the course plans must include greater numbers of incremental steps (and feedback) in the learning process \cite{ale07}, \cite{ss10}, and support personalized learning perhaps via other media such as mobile apps \cite{sg13} to prevent student early frustration and perception of programming as a "difficult task". Devising such interventions can be supported also by qualitative studies on students that believe in a fixed programming aptitude and have low programming-efficacy. 
The authors currently qualitatively study the sub-factors of fixed programming-aptitude belief and low-efficacy.


%



\ifCLASSOPTIONcaptionsoff
  \newpage
\fi



\bibliographystyle{IEEEtran}

\bibliography{mindset}
\end{document}